\documentclass[nofootinbib,aps,showpacs,superscriptaddress,notitlepage]{revtex4-1}

\usepackage{amsmath}
\usepackage{times}
\usepackage{amssymb}
\usepackage{graphics}
\usepackage{epsfig}
\usepackage{bbm}
\usepackage{braket}
\usepackage[latin1]{inputenc} 
\usepackage{color}
\definecolor{shadecolor}{rgb}{0.92,0.92,0.92}
\usepackage{framed}
\usepackage{bbm}

\usepackage[ruled,vlined]{algorithm2e}

\usepackage{geometry}
\usepackage{color}
\definecolor{Ad}{rgb}{0.,0.,0.}
\definecolor{Pr}{rgb}{0.,0.,0.}
\definecolor{Out}{rgb}{0.,0.,0.}

\newcommand{\ad}[1]{{\color{Ad} #1}}
\newcommand{\pr}[1]{{\color{Pr} #1}}

\usepackage[labelfont={bf,small},justification=centerlast,singlelinecheck=false,textfont={small,it}]{caption}
\usepackage{subfig}
\DeclareMathOperator*{\argmin}{arg\,min}
\DeclareMathOperator*{\polylog}{poly\,log}
\newcommand{\tr}{\operatorname{tr}}
\newcommand{\eu}{\mathrm{e}}
\newcommand{\im}{\mathrm{i}\,}
\renewcommand{\Re}{\mathfrak{Re}}
\renewcommand{\Im}{\mathfrak{Im}}

\newcommand{\je}{\textcolor{black}}
\newcommand{\BigO}{O}
\newcommand{\CC}{\mathbb{C}}
\newcommand{\RR}{\mathbb{R}}

\makeatletter
\newcommand{\subalign}[1]{%
	\vcenter{%
		\Let@ \restore@math@cr \default@tag
		\baselineskip\fontdimen10 \scriptfont\tw@
		\advance\baselineskip\fontdimen12 \scriptfont\tw@
		\lineskip\thr@@\fontdimen8 \scriptfont\thr@@
		\lineskiplimit\lineskip
		\ialign{\hfil$\m@th\scriptstyle##$&$\m@th\scriptstyle{}##$\crcr
			#1\crcr
		}%
	}
}
\makeatother

\begin{document}



\title{An efficient quantum algorithm for spectral estimation}


\author{Adrian Steffens}
\affiliation{Dahlem Center for Complex Quantum Systems, Freie Universit\"at Berlin, 14195 Berlin}
\affiliation{Research Laboratory of Electronics,
	Massachusetts Institute of Technology, Cambridge, MA 02139}
\author{Patrick Rebentrost}
\affiliation{Research Laboratory of Electronics,
	Massachusetts Institute of Technology, Cambridge, MA 02139}
\author{Iman Marvian}
\affiliation{Research Laboratory of Electronics,
	Massachusetts Institute of Technology, Cambridge, MA 02139}
\author{Jens Eisert}
\affiliation{Dahlem Center for Complex Quantum Systems, Freie Universit\"at Berlin, 14195 Berlin}
\author{Seth Lloyd}
\affiliation{Research Laboratory of Electronics,
	Massachusetts Institute of Technology, Cambridge, MA 02139}
\affiliation{Department of Mechanical Engineering, Massachusetts Institute of Technology, Cambridge, MA 02139}
\begin{abstract}
We develop an efficient quantum implementation of an important signal processing algorithm for line spectral estimation: the \emph{matrix pencil method}, which determines the frequencies and damping factors of signals consisting of finite sums of exponentially damped sinusoids. Our algorithm provides a quantum speedup in a natural regime where the sampling rate is much higher than the number of sinusoid components. Along the way, we develop techniques that are expected 
to be useful for other quantum algorithms as well---consecutive phase estimations to efficiently make products of asymmetric low rank
matrices classically accessible and an alternative method to efficiently exponentiate non-Hermitian matrices. Our algorithm features an efficient quantum-classical division of labor: The time-critical steps are implemented in quantum superposition, while an interjacent step, requiring only exponentially few parameters, can operate classically. We show that frequencies and damping factors can be obtained in time 
logarithmic in the number of sampling points, exponentially faster than known classical algorithms.
\end{abstract}
\maketitle


\section{Introduction}
Algorithms for the spectral estimation of signals consisting of finite sums of exponentially damped sinosoids have a vast number of practical applications 
in signal processing. These range
from imaging and microscopy~\cite{Karski2009}, radar target identification~\cite{Naishadham2008},  nuclear magnetic resonance spectroscopy~\cite{Viti1997}, estimation of ultra wide-band channels~\cite{Maravic2003}, quantum field tomography~\cite{Steffens2014,cMPSTomographyShort}, 
power electronics~\cite{Leonowicz2003}, up to the simulation of atomic systems~\cite{Andrade2012}.
If the damped frequencies (poles) are known and merely the concomitant coefficients are to be identified, linear methods are readily applicable. In the practically relevant task in which the poles are to be estimated from the data as well, however, one encounters a non-linear problem, and 
significantly more sophisticated methods have to be employed.

There are various so-called high resolution spectral estimation techniques that provide precisely such methods: They include \emph{matrix pencil methods}~\cite{Hua1990}, \emph{Prony's method}~\cite{Prony1795}, \emph{MUSIC}~\cite{Schmidt1986}, \emph{ESPRIT}~\cite{Roy1989}, and 
\emph{atomic norm denoising}~\cite{Bhaskar2013}. 
These techniques are superior to discrete Fourier transform (DFT) in instances with damped signals and close frequencies or small observation time $T>0$~\cite{Park2010,Ri1996,Baqai1993} and are preferred over of the Fourier transform in those applications laid out in Refs.~\cite{Viti1997,Steffens2014,Naishadham2008,Maravic2003,Leonowicz2003,Karski2009,Andrade2012}: The DFT resolution in the frequency domain $\Delta \omega$ is proportional to $1/T$, which is especially critical for poles that are close to each other.
If the poles are sufficiently damped and close, they cannot be resolved by DFT independently of $T$.
Non-linear least-squares fitting of the poles or considering higher-order derivatives of the Fourier transform is in general relatively imprecise, sensitive to noise, or unefficient. 
Non-linear algorithms such as the matrix pencil method can still detect poles, where DFT fails, but are limited to signals composed of finitely many damped sinosoids.

With regard to \emph{quantum algorithms} dedicated to tasks of spectral estimation---algorithms to be run on a quantum computer---the celebrated \emph{quantum Fourier transform} (QFT)~\cite{Nielsen2010} provides an exponential speedup towards the fastest known classical implementations of DFT for processing discretized signals of $N$ samples: Classical fast Fourier transform (FFT) algorithms, on the one hand, take $\Theta(N\log N)$ gates~\cite{Cooley1965}, whereas QFT takes $\Theta(\log^2 N)$ gates to produce a quantum state encoding the Fourier coefficients in its amplitudes. The quantum Fourier transform constitutes a key primitive in various quantum algorithms. In particular, it paved the way for quantum speedups for problems such as prime factoring or order-finding~\cite{Shor1999}.
Regarding spectral estimation, however, QFT inherits the above mentioned properties of its classical counterpart.

The aim of this work is to develop a quantum version of a powerful spectral estimation technique, the matrix pencil method, providing an analogous quantum speedup from $\BigO(\mathrm{poly} \, N)$ to $\BigO(\polylog N)$
for data given in a suitable format. 
Hereto, we make use of the fact that establishing eigenvalues and eigenvectors of low-rank matrices---constituting major steps in this algorithm---can be achieved very fast on quantum computers~\cite{Lloyd2014}. 
Given signal data either via the amplitudes of a quantum state or stored in a \emph{quantum random access memory}~\cite{Giovannetti2008,Giovannetti2008a,DeMartini2009} (QRAM), 
phase estimation of these matrices can be performed directly. For exponentiating non-sparse operators for phase estimation, we employ \emph{quantum principal component analysis} (QPCA) \cite{Lloyd2014} and a recently developed oracle-based method \cite{Rebentrost2016}. 
In an additional step, we employ a quantum linear fitting algorithm~\cite{Wiebe2012,Wang2014} to determine the summing coefficients and hence all parameters that determine the signal function. In this sense, we can understand our algorithm also as an instance of a \emph{non-linear} quantum fitting algorithm in contrast to linear curve fitting algorithms~\cite{Wiebe2012,Wang2014}.
Furthermore, our algorithm can also be employed as a sub-routine in a higher quantum algorithm that requires spectral estimation as an intermediate step.
We expect the developed methods to provide valuable novel primitives to be used in other quantum algorithms as well.

\section{The classical matrix pencil algorithm}\label{sec:MPM}

We start by briefly recapitulating the original (classical) matrix pencil algorithm before in section~\ref{sec:qMPM}, we turn to showing how to 
implement a quantum version of this algorithm in order to gain an exponential speedup. 
Matrix pencil methods (MPM)~\cite{Hua1990} comprise a family of efficient signal processing algorithms for spectral estimation and denoising of equidistantly sampled complex-valued functions $f$ of the type
\begin{equation}\label{eq:1dsignal}
f(t)=\sum_{k=1}^{p}c_{k}\,\eu^{\lambda_{k} t} =:\sum_{k=1}^{p}c_{k}\,\eu^{-\alpha_{k} t}\ \eu^{\im\beta_{k} t},~~~~0\leq t \leq T,
\end{equation}
with the poles $\lambda_k=-\alpha_k+\im\beta_k\in\CC$, damping factors $\alpha_k\in\RR_+$, frequencies $\beta_k\in\RR$, and coefficients $c_k\in\CC$ for $k=1,\dots,p$, where $p\in\mathbb{N}$ is the number of poles. The damping results in a broadening of the spectral lines towards Lorentzian curves. 
Real-valued functions as a special case can be analyzed as well: Here, for each $k=1,\dots,p$ either $\lambda_k,c_k\in\RR$---these terms are non-oscillatory---or 
there exist  $\lambda_{k'},c_{k'}$ such that $\lambda_{k'}=\lambda_k^*$ and $c_{k'}=c_k^*$. Clearly, such signals, in which the 
number of poles $p$ is small and finite, are ubiquitous, or in other instances provide an exceedingly well approximation of the underlying signal.

\begin{shaded}
\begin{algorithm}[H]
  \KwData{Discretized signal with components 
  	$f_{j}=\sum_{k=1}^{p}c_{k}\,\eu^{\lambda_{k}\Delta t\cdot j}$,
 $j=0,\dots,N-1$, $c_k,{\lambda_k\in\CC,}~{\Re(\lambda_k)\leq 0}$.}
  \KwResult{Frequencies $\{\lambda_k\}_{k=1}^p$ and coefficients $\{c_k\}_{k=1}^p$.\\ }
  \caption{Matrix pencil algorithm.}
\end{algorithm}
\end{shaded}


The idea of MPM is to determine the poles $\{\lambda_k\}$ independently from the coefficients $\{c_k\}$ and compare the discretized signal with its translates. Assume that all $c_k$ are nonzero and $\lambda_j\neq\lambda_k$ for $j\neq k$. First, sample the function $f$ equidistantly, 
\begin{equation}
	f\mapsto(f_j)_{j=0}^{N-1},~~~~f_{j}=\sum_{k=1}^{p}c_{k}\,\eu^{\lambda_{k}\Delta t\cdot j},
\end{equation}
with sampling interval $\Delta t>0$.
In general, the higher the number of samples $N$, the more robust the procedure becomes towards noise and the higher the frequencies that can be reconstructed (Nyquist-Shannon sampling theorem~\cite{Shannon1949})---at the expense of computational effort. For clearness, assume that $N$ is even. From the sampled signal, create the 
Hankel matrices $F^{(1)}, F^{(2)}\in \CC^{N/2\,\times\,N/2}$, defined as
\begin{equation}\label{eq:hankel1}
F^{(1)}:=(f_{j+k-2})_{j,k\,=\,1}^{N/2}=
\begin{bmatrix}
f_{0} & f_{1} & \dots & f_{N/2-1}\\
f_{1} & f_{2} & \dots & f_{N/2}\\
\vdots & \vdots &  & \vdots\\
f_{N/2-1} & f_{N/2} & \dots & f_{N-2}
\end{bmatrix}
\end{equation}
and
\begin{equation}
F^{(2)}:=(f_{j+k-1})_{j,k\,=\,1}^{N/2}=
\begin{bmatrix}
f_{1} & f_{2} & \dots & f_{N/2}\\
f_{2} & f_{3} & \dots & f_{N/2+1}\\
\vdots & \vdots &  & \vdots\\
f_{N/2} & f_{N/2+1} & \dots & f_{N-1}
\end{bmatrix} .
\end{equation}
Note that for complex signals, the matrices $F^{(1)}$ and $F^{(2)}$ are symmetric but in general not Hermitian. In other implementations, $F^{(1)}$ and $F^{(2)}$ do not even need to be square. To keep the notation clear, we proceed with square matrices as just defined. Set $\mu_k:=\textrm{e}^{\lambda_{k}\Delta t}$ for $k=1,\dots,p$. It is easy to see that $F^{(1)}$ can be factorized as
\begin{equation}\label{eq:F_decomp1}
	F^{(1)}=M\,D_c\,M^T
\end{equation}
with the Vandermonde matrix $M\in \CC^{N/2\,\times\,p}$,
\begin{equation}
M:=\big(\mu_{k}^{j}\big)_{\subalign{j&=0,\dots,N/2-1\\k&=1,\dots,p}}~=~
\begin{bmatrix}
1 & 1 & \dots & 1\\
\mu_{1} & \mu_{2} & \dots & \mu_{p}\\
\vdots & \vdots &  & \vdots\\
\mu_{1}^{N/2-1} & \mu_{2}^{N/2-1} & \dots & \mu_{p}^{N/2-1}
\end{bmatrix}
\end{equation}
and diagonal matrix $D_c:=\textrm{diag}((c_k)_{k=1}^p)\in\CC^{p\times p}$.
The matrix $F^{(2)}$, on the other hand, can be decomposed as
\begin{equation}\label{eq:F_decomp2}
F^{(2)}=M\,D_c\,D_\mu\,M^T
\end{equation}
with $D_\mu:=\textrm{diag}((\mu_k)_{k=1}^p)\in\CC^{p\times p}$. Note that Eqs.~\eqref{eq:F_decomp1} and~\eqref{eq:F_decomp2} are neither the eigenvalue nor the singular value decomposition of $F^{(1)}$ and $F^{(2)}$, respectively; the column vectors of $M$ do not even have to be orthogonal. We can see from these equations that both $F^{(1)}$ and $F^{(2)}$ have rank $p$, which will in general also be the case for the linear \emph{matrix pencil}~\cite{Golub2012}
\begin{equation}\label{eq:pencil1}
	F^{(2)}-\gamma F^{(1)}=M\,D_c\,(D_\mu-\gamma\mathbbm{1})\,M^T,
\end{equation}
unless $\gamma\in\CC$ matches an element of the set $\{\mu_k\}_{k=1}^p$. Hence, all $\mu_k$ are solutions of the generalized eigenvalue problem (GEVP)
\begin{equation}\label{eq:gen_eigproblem}
	F^{(2)}v=\gamma F^{(1)}v,
\end{equation}
with $v\in\CC^{N/2}$. The matrix pair $(F^{(2)}, F^{(1)})$ is in general \emph{regular} and accordingly results in $N/2$ generalized eigenvalues~\cite{Stewart1990}---not all of these correspond to a $\mu_k$. There are different extensions that take care of this issue and increase algorithmic stability (see, e.g., Ref.~\cite{Hua1991}). To make the algorithm accessible to an efficient quantum implementation, we will consider a specific MPM variant, the \emph{direct} MPM~\cite{Hua1990}:
We make use of the singular value decompositions of $F^{(1)}$ and $F^{(2)}$, keeping only the nonzero singular values and the corresponding singular vectors,
\begin{equation}
 F^{(i)}=U^{(i)}S^{(i)}V^{(i)\,\dagger},~~~~  U^{(i)},V^{(i)}\in\CC^{N/2\,\times\,p}, 
\end{equation} 
with $S^{(i)}\in\CC^{p\times p}$ for $ i=1,2$.
This singular value decomposition of a Hankel matrix of size order $N\times N$ is the time-critical step of the entire algorithm and it scales with $\Theta(N^2\log N)$ using state-of-the-art classical algorithms~~\cite{Browne2009,Xu2008}.
We multiply $U^{(1)\,\dagger}$ from the left and $V^{(1)}$ from the right to
\begin{equation}
F^{(2)}-\gamma F^{(1)}=F^{(2)}-\gamma U^{(1)}S^{(1)}V^{(1)\,\dagger}
\end{equation}
and see that the resulting equivalent GEVP
\begin{equation}\label{eq:pencil2}
U^{(1)\,\dagger}F^{(2)}V^{(1)}\,w=\gamma\, S^{(1)}\,w,
\end{equation}
with $w\in\CC^p$, yields exactly $\{\mu_k\}_{k=1}^p$ as eigenvalues and via $\lambda_k=\log(\mu_k)/\Delta t$ the corresponding poles. The eigenvalues can be retrieved in $\Theta(p^3)$ steps using the QZ algorithm~\cite{Moler1973}.
Although in general it can be numerically favorable to solve the GEVP directly~\cite{Stewart1990}, $S^{(1)}$ is an invertible diagonal matrix and it is in practice sufficient to solve the equivalent ordinary eigenvalue problem
\begin{equation}\label{eq:oEVP}
(S^{(1)})^{-1}U^{(1)\,\dagger}F^{(2)}V^{(1)}\,w=\gamma\, w. 
\end{equation}
The coefficients $\{c_k\}$ are linearly related to the signal and can be obtained by plugging  $\{\mu_k\}_{k=1}^p$ into an overdetermined Vandermonde equation system,
\begin{equation}\label{eq:Vandermonde}
W\,c=\begin{bmatrix}1 & 1 & \cdots & 1\\
\mu_{1} & \mu_{2} & \cdots & \mu_{p}\\
\vdots & \vdots &  & \vdots\\
\mu_{1}^{N-1} & \mu_{1}^{N-1} & \cdots & \mu_p^{N-1}
\end{bmatrix}\cdot
\begin{bmatrix}c_{1}\\
c_{2}\\
\vdots\\
c_p
\end{bmatrix}=
\begin{bmatrix}f_{0}\\
f_{1}\\
\vdots\\
f_{N-1}
\end{bmatrix},
\end{equation}
and computing the least squares solution 
\begin{equation}
\hat{c}:=\argmin_{\tilde{c}\in\CC^p}\Vert W\tilde{c}-f\Vert_2 
\end{equation}
in terms of the $2$-norm,
$\Vert\cdot\Vert_2$, e.g. via applying the Moore-Penrose pseudoinverse $W^+:=(W^\dagger W)^{-1}W^\dagger$ to the signal vector $f$. Thus, all parameters that determine the signal are reconstructed.

\section{Quantum implementation}\label{sec:qMPM}

In the following, we describe how to implement an efficient quantum analogue of the matrix pencil method.
\begin{shaded}
\begin{algorithm}[H]
  \KwData{Discretized signal with components 
  	$f_{j}=\sum_{k=1}^{p}c_{k}\,\eu^{\lambda_{k}\Delta t\cdot j}$,
 $j=0,\dots,N-1$, $c_k,{\lambda_k\in\CC,}~{\Re(\lambda_k)\leq 0}$ either from QRAM or encoded in a quantum state.}
  \KwResult{Frequencies $\{\lambda_k\}_{k=1}^p$ and coefficients $\{c_k\}_{k=1}^p$.\\}
\Begin{
	\begin{itemize}
		\item[] Perform concatenated phase estimations via exponentiating Hermitian matrices $\widetilde{F}^{(1)}, \widetilde{F}^{(2)}$ that contain the matrices $F^{(1)}$, $F^{(2)}$, respectively, 
		yielding the $p$ biggest singular values and the\\ 
		overlaps $\{\langle u^{(1)}_j | u^{(2)}_k\rangle\}$ and $\{\langle v^{(1)}_j | v^{(2)}_k\rangle\}$ of the according left and right singular vectors.

		\item[] Construct the according matrices and solve the eigenvalue problem classically to obtain the poles~$\{\lambda_k\}$.
		\item[] Build a fitting matrix from the poles and obtain the coefficients $\{c_k\}$ via quantum linear fitting. 
	\end{itemize}
}
  \caption{Quantum matrix pencil algorithm.}
\end{algorithm}
\end{shaded}
For an efficient quantum algorithm, we assume that the number of poles $p$ is constant and small relative to the number of samples $N$, which is a natural setting since in practice, we are often interested in damped line spectra with fewer constituents and higher sampling rates for robustness towards noise. The guiding idea is to condense all arrays of size $\BigO(N)$ in Eq.~\eqref{eq:oEVP} into arrays of size $\BigO(p)$ by rewriting the first term in Eq.~\eqref{eq:pencil2},
\begin{eqnarray}\nonumber\!
&&\begingroup\renewcommand*{\arraystretch}{0}
\begin{bmatrix}
\ \rule[2pt]{3em}{.6pt}\ \langle u_{1}^{(1)}|\ \rule[2pt]{3em}{.6pt}\ \\
\raisebox{0.3em}{\vdots\rule[2pt]{6em}{0pt}\vdots}\\
\ \rule[2pt]{3em}{.6pt}\ \langle u_{p}^{(1)}|\ \rule[2pt]{3em}{.6pt}\ 
\end{bmatrix}\endgroup
\!
\begingroup\renewcommand*{\arraycolsep}{0.5pt}\renewcommand*{\arraystretch}{1.2}
\begin{bmatrix}
\mbox{\vline height 3em width .6pt} & \raisebox{1.5em}{\dots}  & \mbox{\vline height 3em width .6pt} \\
|u_{1}^{(2)}\rangle &  & |u_{p}^{(2)}\rangle\\
\mbox{\vline height 3em width .6pt} & \raisebox{1.5em}{\dots}  & \mbox{\vline height 3em width .6pt}
\end{bmatrix}\endgroup
\!
\begingroup\renewcommand*{\arraycolsep}{0.pt}\renewcommand*{\arraystretch}{0}
\begin{bmatrix}
s_{1}^{(2)} &   & 0\\
& \ddots\\
0 &  & s_{p}^{(2)}
\end{bmatrix}\endgroup
\!
\begingroup\renewcommand*{\arraystretch}{0}
\begin{bmatrix}
\ \rule[2pt]{3em}{.6pt}\ \langle v_{1}^{(2)}|\ \rule[2pt]{3em}{.6pt}\ \\
\raisebox{0.3em}{\vdots\rule[2pt]{6em}{0pt}\vdots}\\
\ \rule[2pt]{3em}{.6pt}\ \langle v_{p}^{(2)}|\ \rule[2pt]{3em}{.6pt}\ 
\end{bmatrix}\endgroup
\!
\begingroup\renewcommand*{\arraycolsep}{0.5pt}\renewcommand*{\arraystretch}{1.2}
\begin{bmatrix}
\mbox{\vline height 3em width .6pt} & \raisebox{1.5em}{\dots}  & \mbox{\vline height 3em width .6pt} \\
|v_{1}^{(1)}\rangle &  & |v_{p}^{(1)}\rangle\\
\mbox{\vline height 3em width .6pt} & \raisebox{1.5em}{\dots}  & \mbox{\vline height 3em width .6pt}
\end{bmatrix}\endgroup,
\end{eqnarray}
as
\begin{eqnarray}\label{eq:quantum_pencil}
&&\begin{bmatrix}\langle u_{1}^{(1)}|u_{1}^{(2)}\rangle & \dots & \langle u_{1}^{(1)}|u_{p}^{(2)}\rangle\\
\vdots &  & \vdots\\
\langle u_{p}^{(1)}|u_{1}^{(2)}\rangle & \dots & \langle u_{p}^{(1)}|u_{p}^{(2)}\rangle
\end{bmatrix}\,
\begin{bmatrix}s_{1}^{(2)} &  & 0\\
& \ddots\\
0 &  & s_{p}^{(2)}
\end{bmatrix}\,
\begin{bmatrix}\langle v_{1}^{(2)}|v_{1}^{(1)}\rangle & \dots & \langle v_{1}^{(2)}|v_{p}^{(1)}\rangle\\
\vdots &  & \vdots\\
\langle v_{p}^{(2)}|v_{1}^{(1)}\rangle & \dots & \langle v_{p}^{(2)}|v_{p}^{(1)}\rangle
\end{bmatrix}=:\mathcal{U}\,S^{(2)}\,\mathcal{V},
\end{eqnarray}
with $\mathcal{U},\mathcal{V}\in\CC^{p\times p}$.
The singular values $\{s^{(j)}_k\}$ will be obtained via 
quantum phase estimation~\cite{Kitaev1995,Cleve1998}, the overlaps $\langle v_{k}^{(i)}|v_{l}^{(j)}\rangle$ via two concatenated quantum phase estimations.
The eigenvalue problem Eq.~\eqref{eq:oEVP},
\begin{equation}\label{eq:quantum_pencil2}
(S^{(1)})^{-1}\mathcal{U}\,S^{(2)}\,\mathcal{V}\,w=\gamma\, w,
\end{equation}
is now determined by $2p^2$ complex and $2p$ real numbers, and can easily be evaluated classically in $\Theta(p^3)$ 
operations, yielding the required poles ${\lambda_k=\log(\mu_k)/\Delta t}$ for $k=1,\dots,p$. 
Thus, as other efficient quantum algorithms \cite{Aaronson2009,Rebentrost2014}, the classical result is a low-dimensional read-out quantity. Otherwise, the read-out costs would neutralize any performance gain in the algorithm. 
After that, the poles are used as input for a quantum linear fitting algorithm yielding the coefficients $\{c_k\}$.
In the following, we describe the individual steps of the quantum algorithm in detail. 
We start by discussing the quantum preparation of the Hankel matrices. 


\subsection{\pr{Accessing the data}}

In order to realize a quantum speedup, the signal has to be accessible in a fast and coherent way---otherwise, the read-in process alone would be too costly. 
The data input for the matrix pencil algorithm consists of a time series $(f_j)_{j=0}^{N-1}$.
We consider two crucially different approaches of data access/availability for the quantum algorithm,
with the main focus of this work being on the first approach:
\begin{enumerate}
\item[i)] The signal is stored in a quantum accessible form such as quantum RAM. In other words, we are provided with access to the operation
\begin{equation} \label{eqQRAM}
\ket{j}\ket{0} \mapsto |j \rangle |f_j\rangle
\end{equation}
for $j=0,\dots, N-1$,
with the signal values encoded in binary form in the second quantum register.
In order to create the Hankel matrix $F^{(i)}=(f_{j+k+i-3})_{j,k\,=\,1}^{N/2}\in\mathbb{C}^{N/2\times N/2}$ and $i=1,2$, 
we can perform the following operation with straightforward  index manipulations,
\begin{equation}\label{eq:Hankel_oracle}
\ket{j} \ket{k}\ket{i}\ket{0} \longmapsto \ket{j} \ket{k}\ket{i}\ket{f_{j+k+i-3}}.
\end{equation}
for $j,k=1,\dots,N/2$. The ancilla prepared in $\ket{i}$, $i=1,2$, 
will be used in an entirely classical manner. This operation can be used to simulate Hankel matrices via the non-sparse matrix simulation methods of \cite{Rebentrost2016,Berry2012}. 
One way to implement signal access in Eq.~(\ref{eqQRAM}) is via quantum random access memory (QRAM)~\cite{Giovannetti2008,Giovannetti2008a}.
As \pr{discussed} in Refs.~\cite{Giovannetti2008,Giovannetti2008a}, 
the expected number of hardware elements that are activated in a QRAM call is 
$\BigO(\polylog N)$. 
For each memory call, the amount of required energy and created decoherence thus scales logarithmically with the memory size.
Note that because of their peculiar structure, $(N\times N)$-Hankel matrices 
require only $\BigO (N)$ elements to be stored. In comparison, a general $s$-sparse matrix requires storage of $\BigO (Ns)$ elements. 

\item[ii)] As a second approach, 
we have been given multiple copies of particular quantum state vectors encoding the data  
in their amplitudes. This approach does not require quantum RAM and operates using the  quantum principal component algorithm. Importantly, our method then compares to the Quantum Fourier transform in the sense that it operates on a given initial state that contains the data to be transformed.
The given state vectors have to be of a particular form such as
 \begin{equation}
|\chi^{(i)}\rangle = \frac{1}{\sqrt{C^{(i)}}} \sum_{j,k=1}^{N/2} |j\rangle |k\rangle  \left( F^{(i)}_{j,k}    |0\rangle + 
 a^{(i)} (F^{(i)\dagger} F^{(i)})_{j,k}|1\rangle\right),
 \end{equation}	
 with $C^{(i)}= (\Vert F^{(i)}\Vert_2^2 +a^{(i)\;2} \Vert F^{(i)\dagger} F^{(i)} \Vert_2^2)$ and a known scaling constant $a^{(i)}$ such that $(a^{(i)})^{-1}= O(\max_{j,k} |(F^{(i)\dagger} F^{(i)})_{j,k}|)$, where $\Vert F^{(i)}\Vert_2$ is the Frobenius norm of $F^{(i)}$. 
This state includes in its amplitudes information about the Hankel matrix $F^{(i)}$ and $F^{(i)\dagger} F^{(i)}$.
The particular form of $|\chi^{(i)}\rangle$ will become clear in the next section.  
The advantages of the matrix pencil algorithm over the usual Fourier transform come at a price in the quantum algorithm: We require  
availability of the state vectors
 $|\chi^{(i)}\rangle$  instead of the signal state vector $\sum_j f_{j} \ket{j}$.
\end{enumerate}

 In the next section, we show how the operation in Eq.~(\ref{eqQRAM}) 
 or, alternatively, multiple copies of $|\chi^{(i)}\rangle$ can be used 
 to efficiently simulate a Hermitian matrix that encodes the eigenvalues and associated eigenvectors of the Hankel matrices.

\subsection{Simulating the Hankel matrices}\label{sec:simulating}

We would like to obtain the singular values and vectors of $F^{(1)}$ and $F^{(2)}$ with a quantum speedup via phase estimation, which for real signals correspond, up to signs, to their eigenvalues and vectors. Since the procedure is the same for $F^{(1)}$ and $F^{(2)}$, for clarity we will drop the index in this section and use $F$ for both matrices. Phase estimation requires the repeated application of powers of a unitary operator generated  by  a Hermitian matrix to find the eigenvalues and eigenvectors of that matrix. Thus, we need to connect both Hankel matrices, generally non-Hermitian, to Hermitian matrices. Depending on the input source discussed in the previous section, this is done in different ways. 

Generally, since \ad{$F$ is} not sparse, we cannot make use of the sparse simulation techniques described in Ref.~\cite{Berry2007}. 
Although both matrices have low rank $p\ll N$, they will in general not be positive definite, so that quantum principal component analysis~\cite{Lloyd2014} cannot readily be used either.
Note that although $F^{\dagger}F$ and $FF^{\dagger}$ are positive definite, provide the correct singular vectors of $F$, and 
can be efficiently exponentiated, the phase relations between left and right singular vectors, which are necessary for the matrix pencil algorithm, are not 
preserved. 
This insight can be taken as yet another motivation to look for more general efficient methods to exponentiate
matrices that exhibit a suitable structure, such as being low-rank, sparse or having a low tensor rank. 

For the oracular setting i), we construct a Hermitian matrix $\widetilde{F}$ and apply the unitary operator $\eu^{-\im \widetilde{F}t}$ to an initial quantum state. 
Hereto, we employ the ``extended matrix''
\begin{eqnarray}\label{eqExtendedMatrix}
\widetilde{F}&:=&\left[\begin{array}{cc}
0 & F\\
F^{\dagger} & 0
\end{array}\right]\in\CC^{N\times N},
\end{eqnarray}
which is Hermitian by construction. Its eigenvalues correspond to the singular values $\pm s_{j},j=1,\dots,N/2$, of $F$ and its eigenvectors are proportional to $(u_{j},\pm v_{j})\in\CC^{N}$. 
Importantly, the phase relations between left and right singular vectors are preserved. Note that an operation analogous to Eq.~(\ref{eqQRAM}) for the extended matrix can be easily constructed from Eq.~(\ref{eqQRAM}). 
The method developed in Ref.~\cite{Rebentrost2016} allows us to exponentiate non-sparse Hermitian matrices in this oracular setting.
Following their discussion, Eq.~\eqref{eqExtendedMatrix} is mapped to the corresponding entries of a modified swap matrix $S_{\widetilde{F}}$, which is applied on an initial state $\rho \otimes \sigma$ with auxiliary state 
$\rho:=(1/N)_{j,k=1}^{N}$. This is equivalent to just evolving $\sigma$ in time with the Hamiltonian $\widetilde{F}$ for small ${\Delta t>0}$, i.e.
$\tr_1 ( \eu^{-\im S_{\widetilde{F}}\Delta t}\ \rho\otimes\sigma\ \eu^{\im S_{\widetilde{F}}\Delta t})\approx \eu^{-\im \widetilde{F} \,\Delta t/N}\,\sigma\, \eu^{\im \widetilde{F} \,\Delta t/N}.$
The modified swap matrix $S_{\widetilde{F}}$ is one-sparse within a quadratically larger space and can be efficiently 
exponentiated with the methods in Refs.~\cite{Childs2003, Aharonov2003,Berry2007} 
\ad{with a constant number of oracle calls and run time $\widetilde \BigO(\log N)$, where we omit polylogarithmic factors in $O$ 
by use of the symbol $\widetilde \BigO$.}
Achieving an accuracy $\epsilon>0$ for the eigenvalues requires 
\begin{equation}
\BigO\left ( \frac{ \Vert \widetilde{F} \Vert_{\max}^2 }{\epsilon^3}  \right) 
\end{equation}
steps in the algorithm~\cite{Rebentrost2016}, where $\Vert  \widetilde{F} \Vert_{\max}$ denotes the maximal absolute element of  $\widetilde{F}$.  
Note that in our setting $|\widetilde{F}_{j,k}|=O(1)$ and in particular $\Vert \widetilde{F} \Vert_{\max} =O(1)$. \pr{The run time is the number of steps multiplied by the run time of the swap matrix simulation, i.e.~$\widetilde \BigO\left ( \log N /\varepsilon^3  \right)$.} In Appendix~\ref{sec:Swap_exp} we provide further details on this method. In Appendix \ref{sec:Berry_exp}, we discuss an alternative approach \cite{Berry2012}.

In the setting ii), where we are given multiple copies of state vectors, 
we proceed in a different way and are in a position to employ quantum principal component analysis. The state vector $|\chi\rangle$ can be reduced to a particular quantum density matrix
as
 \begin{equation}
 |\chi\rangle \langle \chi | \longmapsto  \frac{1}{C}\left[
 \begin{array}{cc}
 F F^\dagger & a\, F (F^\dagger F) \\
 a\, (F^\dagger F) F^\dagger & a^2\,(F^\dagger F)(F^\dagger F)
 \end{array}
 \right] =: G.
 \end{equation}
 with quantities $C= (\Vert F \Vert_2^2 +a^{2} \Vert F^{\dagger} F\Vert_2^2)$ and $a^{-1}= O(\max_{j,k} |(F^{\dagger} F)_{j,k}|)$ as before. 
 In the same way, 
  \begin{equation}
 \frac{1}{C}\left[
 \begin{array}{cc}
a^2\,(F^\dagger F)(F^\dagger F)   & a\, F (F^\dagger F) \\
 a\, (F^\dagger F) F^\dagger & F F^\dagger
 \end{array}
 \right] =: \widetilde{G}
 \end{equation}
 can be prepared from a permuted state vector  $\ket{\widetilde\chi}$.
The matrix $Z:=(G+\widetilde{G})/2$ is positive semi-definite with unit trace by construction, just as required by the quantum principal component algorithm. Invoking the singular value decomposition of $F=US V^{\dagger}$, its eigenvalues in terms of the singular values of $F$ are given by 
$s_j^2(a s_j\pm 1)^2/(2C)$,
its eigenvectors are $(u_j,\pm v_j)\in \CC^N$.
The matrix $Z$ has twice the rank of $F$. The application of 
QPCA then allows resolving its eigenvalues to an accuracy $\varepsilon>0$ using
\begin{equation}
\BigO\left ( \frac{1 }{\varepsilon^3}  \right) 
\end{equation}
copies of $|\chi\rangle$ and $\ket{\widetilde{\chi}}$~\cite{Lloyd2014} \pr{for a total run time of again $\widetilde \BigO\left ( \log N/\varepsilon^3  \right)$.} 
In Appendix~\ref{sec:QPCA_exp}, we provide further details on this method.

Both the oracular and the QPCA setting can be employed in quantum phase estimation to obtain the singular values and associated singular vectors of the Hankel matrices in quantum form. Phase estimation allows the preparation of  
\begin{equation}
\sum_{k=1}^{2p}\beta_k \ket{s_k}\ket{u_k,v_k},
\end{equation}
where $F=US V^{\dagger}$ is the singular value decomposition with right and left singular vectors $u_k$ and $v_k$. The associated singular value $s_k$ is encoded in a register. The $\beta_k$ arise from the choice of the initial state. 
The next section describes concretely how consecutive phase estimation steps are used for the matrix pencil algorithm as a building block to obtain the signal poles and expansion coefficients. 

\subsection{Twofold phase estimation}

In this section, we describe how to obtain the singular vector overlaps $\{\mathcal{U}_{j,k}\}$ and $\{\mathcal{V}_{j,k}\}$. Hereto, we perform two concatenated phase estimation procedures to obtain states that encode these overlaps in their amplitudes, which are essentially determined by tomography. It is important to pay attention to the correct phase relations between the overlaps.
Phase estimation is applied to a specific initial state and an additional eigenvalue register. 
\pr{Initial states with large overlap with the eigenstates of $\widetilde F$ or $Z$, respectively, can be prepared efficiently. 
For example, $FF^\dagger /{\rm tr} (F F^\dagger)  \ket{0}\!\bra{0}$ or 
$F^\dagger F /{\rm tr} ( F^\dagger F )  \ket{1}\!\bra{1}$
are suitable initial states and can be prepared from the oracle Eq.~(\ref{eqQRAM}) \cite{Lloyd2014}. For both initial states, the trace with an eigenvector $\ket{u_k,v_k}$ is $\sigma_k^2/(2\sum_j \sigma_j^2)$.
Alternatively, if we have been given multiple copies of $\ket{\chi}$, we can simply take $Z$ to be the initial state \cite{Lloyd2014}.
}

We append two registers for storing the singular values to the initial state, obtaining $\ket{0}\ket{0}\ket{\psi_0}$ with the notation 
$\ket{0}:=\ket{0, \hdots , 0}$, and perform the phase estimation algorithm with $\textrm{e}^{-\im S_{\tilde F^{(2)}}\,\Delta t}$ as a unitary operator to obtain a state proportional to
\begin{equation}
\sum_{k=1}^{2p}\braket{u_k^{(2)},v_k^{(2)}|\psi_0}\ket{0}\ket{s_k^{(2)}}\ket{u_k^{(2)},v_k^{(2)}},
\end{equation}
where for clarity we order the eigenspaces such that positive singular values are put first, i.e. $s_{k+p}^{(2)}=-s_k^{(2)}$, $u^{(2)}_{k+p}=u^{(2)}_{k}$, and $v^{(2)}_{k+p}=-v^{(2)}_{k}$ for $k=1,\dots,p$. To obtain the overlaps of the matrices $U^{(1)}$ and $U^{(2)}$, the $v$-part of the eigenvector of $\widetilde{F}^{(2)}$ is projected out, yielding
\begin{widetext} 
\begin{equation}\label{eq:gk}
\ket{\psi_1}=\frac{1}{\nu_1}\sum_{k=1}^{2p}\braket{u_k^{(2)},v_k^{(2)}|\psi_0}\ket{0}\ket{s_k^{(2)}}\ket{u_k^{(2)},0}=:\sum_{j=1}^{2p} g_{k}\ket{0}\ket{s_k^{(2)}}\ket{u_k^{(2)},0}
\end{equation}
\end{widetext} 
with normalization factor $\nu_1\in\RR_+$ and $\sum_{k=1}^{2p}\vert g_k\vert^2=1$. Each singular value $s_k^{(2)}\in\RR_+$ can be determined efficiently from this with accuracy $\epsilon_\sigma$ in a runtime of $ \tilde \BigO(\log N/\epsilon_\sigma^3)$ (cf. Sec.~\ref{sec:simulating}). 
We need to determine the amplitudes $\{g_k\}$, which have to be removed from the overlap values. For this, we essentially perform standard tomography of the quantum state Eq.~\eqref{eq:gk}.
The singular register vectors $\{\ket{s_k^{(2)}}\}_{k=1}^{2p}$ are pairwise orthogonal,  
so that the amplitudes $\{g_k\}_{k=1}^p$ can be efficiently obtained---up to a global complex phase $\eu^{\im\vartheta_1}$---via measurements e.g. of the form 
\begin{equation}
\ket{s^{(2)}_{k_1}}\bra{s^{(2)}_{k_1}},~~~\ket{s^{(2)}_{k_2}}\bra{s^{(2)}_{k_2}},~~~(\ket{s^{(2)}_{k_1}}+\ket{s^{(2)}_{k_2}})(\bra{s^{(2)}_{k_1}}+\bra{s^{(2)}_{k_2}}),
~~~(\ket{s^{(2)}_{k_1}}-\mathrm{i}\ket{s^{(2)}_{k_2}})(\bra{s^{(2)}_{k_1}}+\mathrm{i}\bra{s^{(2)}_{k_2}}),
\end{equation}
with probabilities 
\begin{equation}\label{eq:probabilities_g}
\vert g_{k_1}\vert^2,~~~\vert g_{k_2}\vert^2,~~~\vert g_{k_1}\vert^2+\vert g_{k_2}\vert^2+2\,\Re(g_{k_1}g_{k_2}^*),~~~\vert g_{k_1}\vert^2+\vert g_{k_2}\vert^2+2\,\Im(g_{k_1}g_{k_2}^*),
\end{equation}
respectively. Suppose $g_{k_1}$ is known. Then $g_{k_2}$ can easily be obtained from Eq.~\eqref{eq:probabilities_g}. Hence, by fixing one global phase $\eu^{\im\vartheta_1}$ (e.g. corresponding to $g_1\overset{!}{=}+|g_1|\,$), all values $\{g_k\}_{k=1}^{2p}$ are unambiguously determined. 
Requiring an accuracy 
\begin{equation}
	\varepsilon_g= {\mathbb{V}(g)}^{1/2}/\mathbb{E}(g)
\end{equation}
of the probabilities in Eq.~\eqref{eq:probabilities_g} for $k=1,\dots,p$, we require 
$\BigO(\xi_g/ \varepsilon_g^2)$ measurement repetitions for each amplitude, 
denoting the ratio of the biggest over the smallest probability with $\xi_g$. 
We thus have established the values 
\begin{equation}\label{eq:g_k}
g_k\eu^{\im\vartheta_1}=\braket{u_k^{(2)},v_k^{(2)}|\psi_0}\frac{\eu^{\im\vartheta_1}}{\nu_1},~~~k=1,\dots,2p.
\end{equation}
Next, the state vector 
$\ket{\psi_1}$ is used as input for a second phase estimation procedure with  $\eu^{-\im S_{\widetilde{F}^{(1)}}\Delta t}$ as unitary operator, yielding
\begin{align}
\ket{\psi_2}=\frac{1}{\nu_2}&\sum_{j,k=1}^{2p} \braket{u_k^{(2)},v_k^{(2)}|\psi_0}\braket{u_j^{(1)},v_j^{(1)}|u_k^{(2)},0}\ket{s_j^{(1)}}
\ket{s_k^{(2)}}\ket{u_j^{(1)},v_k^{(1)}}\nonumber\\
=:&\sum_{j,k=1}^{2p} h_{j,k}\ket{s_j^{(1)}}
\ket{s_k^{(2)}}\ket{u_j^{(1)},v_k^{(1)}}
\end{align}
with normalization factor $\nu_2\in\RR_+$ and $\sum_{j,k=1}^{2p}\vert h_{j,k}\vert^2=1$. The inner product $\braket{u_j^{(1)},v_k^{(1)}|u_k^{(2)},0}$ reduces to $\braket{u_j^{(1)}|u_k^{(2)}}$ with vectors in $\CC^N$. The same way as above, we determine the singular values $\{s_j^{(1)}\}$ and the values 
\begin{equation}\label{eq:h_jk}
h_{j,k}\eu^{\im\vartheta_2}=\braket{u_k^{(2)},v_k^{(2)}|\psi_0}\braket{u_j^{(1)}|u_k^{(2)}}\frac{\eu^{\im\vartheta_2}}{\nu_2},~~~j,k=1,\dots,2p,
\end{equation}
up to $\varepsilon_h$ with global phase $\eu^{\im\vartheta_2}$ \pr {with} $\BigO( \xi_h/ {\varepsilon_h^2})$ \pr{repetitions} for each amplitude. 
Dividing the values in Eq.~\eqref{eq:h_jk} by the ones in Eq.~\eqref{eq:g_k}, we obtain
\begin{equation}\label{eq:U_jk}
\mathcal{U}_{j,k}\,\nu_\mathcal{U}\,\eu^{\im\vartheta_\mathcal{U}}=\braket{u_j^{(1)}|u_k^{(2)}}\nu_\mathcal{U}\,\eu^{\im\vartheta_\mathcal{U}},~~~j,k=1,\dots,2p,
\end{equation}
with $\vartheta_\mathcal{U}:=\vartheta_2-\vartheta_1$, $\nu_\mathcal{U}:=\nu_1/\nu_2$ and accuracy $\sim \varepsilon_g+\varepsilon_h$. 
The established overlaps
\begin{equation}
\braket{u_j^{(1)}|u_k^{(2)}},~\braket{u_{j+p}^{(1)}|u_k^{(2)}},~\braket{u_j^{(1)}|u_{k+p}^{(2)}},~\braket{u_{j+p}^{(1)}|u_{k+p}^{(2)}}
\end{equation}
correspond to the same matrix entry of $\mathcal{U}$ for $j,k=1,\dots,p$ and can be averaged over.
This way, the matrix $\mathcal{U}$ is determined up to a global phase and a normalization factor.
Repeating the entire procedure, but with projecting out the $u$-part,
\begin{equation}
\ket{u_k^{(2)},v_k^{(2)}} \mapsto \ket{0,v_k^{(2)}},~~k=1,\dots,2p,
\end{equation}
yields all overlaps $\{\braket{v_{j}^{(1)}|v_k^{(2)}}\}_{j,k=1}^p$, the entries of $\mathcal{V}$, up to a factor $\nu_\mathcal{V}\,\textrm{e}^{\textrm{i}\vartheta_\mathcal{V}}$. Note that
\begin{equation}
\braket{v_j^{(1)}|v_k^{(2)}}=-\braket{v_{j+p}^{(1)}|v_k^{(2)}}=-\braket{v_j^{(1)}|v_{k+p}^{(2)}}=\braket{v_{j+p}^{(1)}|v_{k+p}^{(2)}}
\end{equation}
for $j,k=1,\dots,p$ because the $v$-parts of the $\widetilde{F}^{(i)}$ eigenvectors from $k=1,\dots,p$ and $k=p+1,\dots,2p$ have opposite signs. For real-valued signals and Hermitian $F^{(i)}$, we can perform the procedure with $\textrm{e}^{-\im S_{F^{(i)}}\,\Delta t}$ instead of $\textrm{e}^{-\im S_{\widetilde{F}^{(i)}}\,\Delta t}$ and do not need to project the $u$- and $v$-parts.
 
In summary, we have determined the singular values forming matrix $S^{(i)}$ to accuracy $\epsilon_\sigma$ in time $ \widetilde \BigO(p/\epsilon_\sigma^3)$. In addition,  we have determined the overlaps of the right and left singular vectors of the two Hankel matrices $F^{(1)}$ and $F^{(2)}$. The required 
number of repetitions is  
\begin{equation}
n_\mathcal{U}=\BigO\left(\frac{p}{\varepsilon_g^2}\xi_g+\frac{p^2}{\varepsilon_h^2}\xi_h\right)
\end{equation}
for obtaining the entries of $\mathcal{U}$ and
analogously $n_\mathcal{V}$ for obtaining the entries of $\mathcal{V}$.
With 
\begin{equation}
\pr{	n_{\phi}=\widetilde \BigO\left(\frac{\log N}{\varepsilon^3}\right)}
\end{equation}	
for the cost of the phase estimation, this leads to a total \pr{run time} of 
\begin{equation}
\pr{n:=n_\phi \,(n_\mathcal{U}+n_\mathcal{V})\,=\widetilde \BigO\left(\frac{p^2\, \xi\, }{\varepsilon^5}\log N\right ),}
\end{equation}
with $\xi:=\max\, \{\xi_g,\xi_h\}$. The performance scales as $n=\BigO ({\rm poly} \log N)$ for example in the following regime: First, the number of poles is small compared to $N$, which is a natural regime, as mentioned above; second, regarding $\xi$, if the overlaps are of the same order of magnitude, $\xi= \BigO ({\rm poly} \log N)$; and third, an error $1/\varepsilon=\BigO ({\rm poly} \log N)$ can be tolerated.
 
\subsection{Solving the small classical problem} 
Having determined the values via phase estimation, the reconstructed eigenvalue equation~\eqref{eq:quantum_pencil2} now reads
\begin{equation}\label{eq:quantum_pencil3}
\hat{\mathcal{F}}\,w:=\nu_\mathcal{U}\nu_\mathcal{V}\textrm{e}^{\textrm{i}(\vartheta_\mathcal{U}+\vartheta_\mathcal{V})}\ (S^{(1)})^{-1}\mathcal{U}\,S^{(2)}\,\mathcal{V}\,w=\gamma \,w.
\end{equation}
All (scaled) matrix entries of Eq.~\eqref{eq:quantum_pencil3} are available classically and we can solve the problem with a classical algorithm~\cite{Moler1973} running with time $\BigO(p^3)$. The errors in the matrix entries are amplified within the entries of the matrix product entries $\hat{\mathcal{F}}_{j,k}$ by a factor of $\mathrm{poly}\ p$ at worst. 
Taking the inverse of $S^{(1)}$ amounts to inverting its diagonal entries, hence the relative errors of $(S^{(1)})^{-1}_{j,j}$ are unchanged. These are only small if the effective singular values of $F^{(1)}$ (the ones bigger than a suitable threshold $\theta_1$) are sufficiently bigger than zero, resulting in a condition number of $S^{(1)}$ bounded by $\max_j (S^{(1)}_{j,j})/\theta_1$.
$\mathcal{F}$ as well as the perturbed matrix $\hat{\mathcal{F}}=\mathcal{F}+\Delta \mathcal{F}$ will in general not be normal, but diagonalizable: $\mathcal{F}=X\,\mathrm{diag}(\lambda_j)\,X^{-1}$. According to the Bauer-Fike theorem~\cite{Bauer1960}, we can order the eigenvalues $\{\hat{\lambda}_j\}$ of $\hat{\mathcal{F}}$ such that
\begin{equation}\label{eq:bauerfike}
|\lambda_j-\hat{\lambda}_j|\leq \kappa(X)\Vert \mathcal{F}-\hat{\mathcal{F}}\Vert_2
\end{equation}
for $j=1,\dots,p$, where $\kappa(X):=\Vert X \Vert_2 \Vert X^{-1}\Vert_2$ is the condition number of $X$, which represents the amplification factor of the matrix perturbation towards the perturbation of the eigenvalues. 
The matrix perturbation contributes linearly, while the condition number of $X$, which is independent of the perturbation $\Delta \mathcal{F}$, is related to the condition of the underlying inverse spectral estimation problem. This could in principle be ill-conditioned (e.g. for the reconstruction of extremely small or highly damped spectral components relative to the other ones), but we are more concerned with problems that are also of interest in the classical world and hence sufficiently well-conditioned.
Note that $p$, the number of poles, is small by assumption so that this classical step does not pose a computational bottleneck for the algorithm.  
For noisy signals, the rank of $F^{(i)}$ will in general be larger than $p$, $F^{(i)}$ could even be full rank---for not too large noise, however, the additional noise components will remain small such that the effective rank will still be at $p$. Since only the biggest components of $F^{(i)}$ are taken into account, this results in a rank-$p$ approximation that is best in the Frobenius norm sense (Eckart-Young theorem~\cite{Eckart1936}) and an effective noise filtering of the underlying signal.

The eigenvalues $\gamma_k$ of Eq.~\eqref{eq:quantum_pencil3} are determined up to $\textrm{e}^{-\textrm{i}(\varphi_\mathcal{U}+\varphi_\mathcal{V})-\log(\nu_\mathcal{U}\nu_\mathcal{V})}$, which corresponds to a uniform translation of all poles. We can take care of this ambiguity by introducing an additional reference pole $\lambda_\textrm{ref}:=0$ (corresponding to the eigenvalue $\mu_\mathrm{ref}=1$) that has to be incorporated into the original signal. This can easily be achieved by adding any constant to the original signal vector (its normalizability is not affected). Since for exponentially damped signals  $\mathfrak{Re}(\lambda_k)\leq 0$ holds for each $k$, the eigenvalue $\gamma_\mathrm{ref}$ corresponding to the reference pole will still be identifiable as the one with the biggest absolute value $|\gamma_k|$.  Simply dividing all $\gamma_k$ by $\gamma_\mathrm{ref}$ (corresponding to the transformation $\lambda_k\Delta t\mapsto\lambda_k\Delta t+\textrm{i}(\varphi_\mathcal{U}+\varphi_\mathcal{V})+\log(\nu_\mathcal{U}\nu_\mathcal{V})$ for each $k$) then yields the correct values $\{\mu_k\}$ and poles.

\subsection{Quantum linear fitting}
We feed the poles back into the quantum world by using the quantum fitting algorithm described in Refs.~\cite{Wiebe2012,Wang2014} to obtain the coefficients $\{c_k\}$ in $\BigO(\log(N)p)$ steps and hence the entire parametrization of the input function. We consider real and imaginary parts of the signal $f$, the poles $\lambda_k\,\Delta t=:-\alpha_k+\im \beta_k$ and the coefficients $c_k=a_k+\im b_k$ separately, and Eq.~\eqref{eq:Vandermonde} becomes
\begin{widetext}
\begin{equation}\label{eq:linfit_real}
\widetilde{W}\,\tilde{c}=\tilde{f}
\end{equation}
with
\begin{equation}\nonumber
\widetilde{W}:=\begin{bmatrix}
\eu^{-\alpha_1 \cdot 0}\cos(\beta_1\cdot 0) & \dots & \eu^{-\alpha_p \cdot 0}\cos(\beta_p\cdot 0) &
-\eu^{-\alpha_1 \cdot 0}\sin(\beta_1\cdot 0) & \dots & -\eu^{-\alpha_p \cdot 0}\sin(\beta_p\cdot 0)\\
\vdots& &\vdots&\vdots& &\vdots\\
\eu^{-\alpha_1\cdot\widetilde{N}}\cos(\beta_1\!\cdot\!\widetilde{N}) & \dots & \eu^{-\alpha_p\cdot\widetilde{N}}\cos(\beta_p\!\cdot\!\widetilde{N} )&
-\eu^{-\alpha_1\cdot\widetilde{N}}\sin(\beta_1\!\cdot\!\widetilde{N} )& \dots & -\eu^{-\alpha_p\cdot\widetilde{N}}\sin(\beta_p\!\cdot\!\widetilde{N})\\
\eu^{-\alpha_1 \cdot 0}\sin(\beta_1\cdot 0) & \dots & \eu^{-\alpha_p \cdot 0}\sin(\beta_p\cdot 0) &
\eu^{-\alpha_1 \cdot 0}\cos(\beta_1\cdot 0) & \dots & \eu^{-\alpha_p \cdot 0}\cos(\beta_p\cdot 0)\\
\vdots& &\vdots&\vdots& &\vdots\\
\eu^{-\alpha_1 \cdot \widetilde{N}}\sin(\beta_1\!\cdot\! \widetilde{N}) & \dots & \eu^{-\alpha_p \cdot \widetilde{N}}\sin(\beta_p\!\cdot\! \widetilde{N}) &
\eu^{-\alpha_1 \cdot \widetilde{N}}\cos(\beta_1\!\cdot\! \widetilde{N}) & \dots & \eu^{-\alpha_p \cdot \widetilde{N}}\cos(\beta_p\!\cdot\!\widetilde{N})
\end{bmatrix},
\end{equation}
\begin{equation}\nonumber
\widetilde{W}:=(w_{j,k})=
\begin{bmatrix}
(\Re\,\mu_k^{j-1}) & (-\Im\,\mu_k^{j-1})\\
 (\Im\,\mu_k^{j-1})& (\Re\,\mu_k^{j-1})
\end{bmatrix}\in\RR^{2N\times 2p},~~~
\tilde{c}:=\begin{bmatrix}
\Re\,c_1\\ \vdots\\\Re\,c_p\\\Im\,c_1\\ \vdots\\\Im\,c_p
\end{bmatrix}\in\RR^{2p},~~~
\tilde{f}:=\begin{bmatrix}
\Re\,f_0\\ \vdots\\\Re\,f_{\widetilde{N}}\\\Im\,f_0\\ \vdots\\\Im\,f_{\widetilde{N}}
\end{bmatrix}\in\RR^{2N},
\end{equation}
and $\widetilde{N}:=N-1$.
The vector $2$-norm of the $k$-th column of $\widetilde{W}$ can be established in closed form as
\begin{equation}
\frac{1-\eu^{-2\alpha_k N}}{1-\eu^{-2\alpha_k}} \textrm{,~if~}\alpha_k>0, ~~~\textrm{and~~} N \textrm{,~if~}\alpha_k=0.
\end{equation}
Hence, $\Vert\widetilde{W}\Vert_2$ can be computed in time $\BigO(p)$. We will rescale the solution for $c$ such that we can assume that $\Vert\widetilde{W}\Vert_2=1$. The norms of matrices $\Vert\widetilde{W}\Vert_2$ for real-valued signals can be calculated as well by combining the norms of the $k$-th with the $(k+p)$-th column. Since each row consists of $2p$ elements, the row norms can be computed in $\BigO(p)$ as well. 
\end{widetext}

Since $\alpha:=(\alpha_k),\beta:=(\beta_k)$ are known, we can construct a quantum oracle, providing quantum access to the matrix entries $w_{j,k}(\alpha,\beta)$,
\begin{equation}
\ket{\alpha}\ket{\beta}\ket{j}\ket{k}\ket{0}\,\longmapsto\,\ket{\alpha}\ket{\beta}\ket{j}\ket{k}\ket{w_{j,k}(\alpha,\beta)}\je{.}
\end{equation}
The matrix $\widetilde{W}$ can be prepared as a state vector
\begin{equation}
\ket{w}=\sum_{j=1}^{2N}\sum_{k=1}^{2p}w_{j,k}\ket{j}\ket{k}
\end{equation}
following the procedure described in Ref.~\cite{Wang2014}
with time $\widetilde{\BigO}(\polylog(N)\,p\, \xi_W \log(1/\zeta))$, where $\zeta$ is the accuracy of the preparation of $\ket{w}$ and 
\begin{equation}
\xi_W:=\max\Vert w_j\Vert_2/\min\Vert w_j\Vert_2. 
\end{equation}
Here, we set $\widetilde{\BigO}(g(N)):=\BigO(g(N) \polylog (g(N)))$ for functions $g$. For the preparation of $\ket{\tilde{f}}$, we require time $\widetilde{\BigO}(\polylog(N)\, \xi_f \log(1/\zeta))$ with 
 \begin{equation}
 	\xi_f:=\max\vert \tilde{f}_j\vert_{}/\min\vert \tilde{f}_j\vert.
\end{equation}
With $\ket{w}$ and $\ket{f}$ prepared, we then can proceed as described in Ref.~\cite[Th.~2,3]{Wang2014} and obtain with probability bigger than $2/3$ an estimate $\hat{c}$ in time 
$\allowbreak\widetilde{\BigO} (\polylog(N)\kappa_W p^{3/2}(\sqrt{2p}\xi_f/\varepsilon+\kappa_W^2\xi_f/\Phi+\kappa_W^6(2p)^5\xi_W/\varepsilon^4\Phi)/\\\varepsilon\Phi)$, with 2-norm accuracy $\varepsilon$,   $\kappa_W=\Vert\widetilde{W}\Vert_2/\Vert\widetilde{W}^+\Vert_2$, and norm $\Phi$ of the projection of $\tilde{f}$ onto the column space of $\widetilde{W}$, the fit quality.
Importantly, we can estimate the quality of the fit with time $\widetilde{\BigO}(\polylog(N)(\xi_f+\xi_W(2p)^3\kappa_W^4/\varepsilon)/\varepsilon)$.
Note that sampling $\hat{c}$ is efficient because it comprises $\BigO(p)$ components. 
Altogether, we \je{have} determined the sought-after coefficients and hence all parameters that characterize the signal $f$ in $\polylog N$. This concludes the description of the quantum matrix pencil algorithm.

\section{Summary and discussion}

We have developed a quantum implementation of an important algorithm for spectral estimation, the matrix pencil method, 
taking a tool from signal processing to the quantum world \je{and significantly improving upon the effort required}. 
\ad{Given the arguable scarcity of quantum algorithms with this feature, progress in this respect seems highly desirable.}
The quantum MPM is a useful alternative to quantum Fourier transform in many practical applications, in the same way that classical MPMs and related algorithms are useful alternatives to the classical Fourier transform. This is especially the case for signals with close damped poles and limited total sampling time. 

For a signal given by $N$ equidistant samples, we have made use of the fact that the eigenvalue problem Eq.~\eqref{eq:quantum_pencil2} consisted of large matrices of size $\BigO(N)$ that could, 
however, be contracted into manageable matrices of size $\BigO(p)$ via concatenated use quantum phase estimations in $\BigO(\polylog N)$. 
This justifies the use of a quantum version of the matrix pencil method as opposed to quantum versions of related algorithms like Prony's method, where the $p$ quantities leading the corresponding poles are determined in a later step, during the fitting of the coefficients, and the critical step would already be $\BigO(\mathrm{poly} N)$.

The quantum phase estimation was shown to be implementable in two complementary ways: either by retrieving the input signal via quantum oracle calls such as QRAM, or by using multiple copies of a state with the signal encoded in its amplitudes for quantum principal component analysis. The employed extended matrix construction allows for exponentiating more general matrices via QPCA that are not positive semidefinite. This provides a useful new primitive also for other quantum algorithms.

The actual step to determine the poles from an eigenvalue problem of a $p\times p$ matrix \ad{can be performed} classically since $p$ is assumed to be small. Subsequently, feeding back the established poles into a quantum fitting algorithm allows the coefficients of the signal \ad{again to be} determined efficiently in $\BigO(\polylog N)$. This way, we have an effective division of labor between classical and quantum algorithms, to the extent that such a hybrid algorithm is possible efficiently. Classical intermediate steps are for example reminiscent of quantum error correction, where error syndromes are measured and the quantum state is processed according to the classical measurement results \cite{Gottesman2009}. 


The outlined procedure is generalizable to arbitrary signal dimensions $d$, i.e. signals of the type
$f(t_1,\dots,t_d) = \sum_{k_{1},\dots,k_{d}=1}^{p} c_{k_{1},\dots,k_d} \,\eu^{\lambda_{k_{1}}t_{1}+\dots+\lambda_{k_d}t_d}$, with $c\in{\CC^p}^{\,d}$ by suitable tensor contractions of the array of signal samples $(f_{j_1,\dots,j_d})_{\{j_l\}=0}^{N-1}$~\cite{Steffens2014} or fixing all time indices but one and applying the MPM on the remaining vector. This yields the sought-after poles since they are the same for the different time indices $t_i$. For time index-dependent poles, one can consider ``enhanced matrices''---embeddings of Hankel matrices that correspond to one-dimensional projections of the multidimensional signal within a larger block Hankel matrix---as in Ref.~\cite{Hua1992}. There are many potential applications for this, e.g. in radar imaging and geophysics~\cite{Garello2013}.
\je{We expect the methods and primitives that we develop and introduce here to be highly 
useful also when devising other quantum algorithms. This includes the computation of overlaps by suitably concatenating quantum phase
estimation procedures and the efficient exponentiation of structured matrices on a quantum computer. We hope that the present work
stimulates such further research.}

\section*{Acknowledgments}

AS thanks the German National Academic Foundation (Studienstiftung des deutschen Volkes) and the Fritz Haber Institute of the Max Planck Society for support. SL and PR were supported by ARO and AFOSR.
JE thanks the Templeton Foundation, the DFG (CRC 183, EI 519/7-1), the ERC (TAQ), and the EC (RAQUEL, AQuS) for support.

\bibliographystyle{unsrt}

\appendix

\section{Non-sparse oracle method via modified swap matrix}\label{sec:Swap_exp}
A new method developed in Ref.~\cite{Rebentrost2016} allows us to exponentiate indefinite Hermitian matrices. 
In appendices~\ref{sec:QPCA_exp} and \ref{sec:Berry_exp}, 
we discuss alternative ways forward, contributing to providing 
a wider framework of efficient matrix exponentiation.
Following the discussion in Ref.~\cite{Rebentrost2016}, Eq.~\eqref{eq:Hankel_oracle} is mapped to the corresponding entry of a modified swap matrix $S_{\widetilde{F}^{(i)}}$, resulting in the matrix
\begin{equation}
S_{\widetilde{F}^{(i)}}:=\sum_{j,k=1}^{N}\widetilde{F}^{(i)}_{j,k}\ket{k}\!\bra{j}\otimes\ket{j}\!\bra{k}\ \in\ \CC^{\,N^2 \times N^2}.
\end{equation}
In  \cite{Rebentrost2016} it is
 shown that performing infinitesimal swap operations on an initial state $\rho \otimes \sigma$ with auxiliary state 
$\rho:=(1/N)_{j,k=1}^{N}$ is equivalent to just evolving $\sigma$ in time with the Hamiltonian $\widetilde{F}^{(i)}$ for small ${\Delta t>0}$, i.e.
$\tr_1 ( \eu^{-\im S_{\widetilde{F}^{(i)}}\Delta t}\ \rho\otimes\sigma\ \eu^{\im S_{\widetilde{F}^{(i)}}\Delta t})\approx \eu^{-\im \widetilde{F}^{(i)} \,\Delta t/N}\,\sigma\, \eu^{\im \widetilde{F}^{(i)} \,\Delta t/N}.$


\section{Alternative non-sparse quantum oracle method}\label{sec:Berry_exp}

Berry et al. present a method to exponentiate matrices sublinear in the sparsity \cite{Berry2012}. 
In this section, we summarize the performance and requirements of this method and the application to the low-rank Hankel matrices of the present work. 
The number of oracle queries for simulating a matrix such as the Hermitian $\widetilde{F}^{(i)}$ in Eq.~(\ref{eqExtendedMatrix}) is given by 
\begin{equation}O(t^{3/2} \sqrt{{s \Lambda_{\rm tot}}/{\epsilon}}),
\end{equation}	
where $s$ is the sparsity and $\epsilon$ is the error.
The quantity $\Lambda_{\rm tot}>0$ 
depends on the norms of the matrix as $\Lambda_{\rm tot} = \Lambda \Lambda_1 \Lambda_{\max}$ with the spectral norm $\Lambda= \Vert \widetilde{F}^{(i)} \Vert_\infty$, the maximum column sum norm $\Lambda_1 = \Vert \widetilde{F}^{(i)} \Vert_1$, and the maximum matrix element $\Lambda_{\max} = \Vert \widetilde{F}^{(i)} \Vert_{\max}$. The conditions for this to work are given by $\Lambda t \geq \sqrt{\epsilon}$,
\begin{equation}
	 t \geq \frac{\Lambda}{\Lambda_{\max} \Lambda_1 s}, 
\end{equation}	
and $\Lambda \leq \Lambda_1$.

We confirm that under reasonable assumptions the low-rank non-sparse Hankel matrices under consideration in this work can be simulated with $O(\log N)$ queries. Assume that the signal is reasonably small with not too many zeros. This implies that the matrix $\widetilde{F}^{(i)}$ is non-sparse with $s=\Theta(N)$ and the individual elements scale as $\widetilde{F}^{(i)}_{jk}=\Theta(1)$. If we assume that the signal is generated by a few (in fact, $p$) components, then the matrix is low rank with rank $2p$. Since ${\rm tr ( (\widetilde{F}^{(i)})^2 ) }=\sum_{j=1}^{2p} \lambda^2_j \leq N^2 \Vert \widetilde{F}^{(i)} \Vert_{\max}^2$, we have that the significant eigenvalues scale as $\lambda_j=\Theta(N)$, $j=1, \dots, 2p$.  These assumptions have the following straightforward implications:
\begin{enumerate}
\item[i)] The spectral norm (largest eigenvalue) is $\Lambda=\Theta(N)$, \item[ii)] the induced $1$-norm (maximum column sum) is $\Lambda_1=\Theta (N)$, and 
\item [iii)] the maximum element is $\Lambda_{\max} = \Theta(1)$. 
\end{enumerate}
Thus, $\Lambda_{\rm tot} = \Theta(N^2)$ and the total number of queries is
$O(t^{3/2} \sqrt{{\Theta(N^3)}/{\epsilon}})$. We need time $t=\Theta(1/N)$ to resolve the eigenvalues $\lambda_j=\Theta(N)$ via phase estimation. Thus, at an error $\epsilon$, we need $O(1/\sqrt{\epsilon})$ queries, which is again efficient. 

We show that we can satisfy the conditions as follows. Since we have $t=\Theta (1/N)$ already from phase estimation, we can assume that with constant effort  $  t \geq \sqrt{\epsilon}/\Lambda=\Theta(\sqrt{\epsilon}/N)$. Next, by using i)-iii) and $s=\Theta(N)$, we have
\begin{equation}
	 t \geq \frac{\Lambda}{\Lambda_{\max} \Lambda_1 s} =\Theta\
\left(\frac{1}{N}\right).
	\end{equation} 
	  The third criterion $\Lambda \leq \Lambda_1$ is satisfied by Gershgorin's theorem, since the eigenvalues are bounded by the maximum sum of the absolute elements in a row/column.

\section{Matrix exponentiation via quantum principal component analysis}
\label{sec:QPCA_exp}

In this appendix, we present an alternative way to efficiently exponentiate indefinite matrices, in order
to give more substance to ideas of exponentiating structured matrices while at the same time 
preserving a phase relationship. Since exponentiating matrices $F\in \CC^{N/2\times N/2}$ 
while a preserving phase relationship is key to the above algorithm and is expected to be important in other quantum algorithms, we briefly
present an alternative method that accomplishes this task via quantum principal component analysis. 
This method compares to the quantum Fourier transform in the sense that it operates on a given initial state that contains the data to be transformed in its amplitudes. We assume that we have been presented with many copies of the state vector
 \begin{equation}
|\chi\rangle = \frac{1}{\sqrt{C}} \sum_{j,k=1}^{N/2} |j\rangle |k\rangle  \left( F_{j,k}  |0\rangle + 
 a (F^\dagger F)_{j,k}  |1\rangle\right),
 \end{equation}	
 with $C:= (\Vert F\Vert_2^2 +a^2 \Vert F^\dagger F \Vert_2^2)$ and $a^{-1}:= O(\max_{j,k} |(F^\dagger F)_{j,k}|)$. The matrix $F$ takes the role of $F^{(1)}$ and $F^{(2)}$ of the main text, so again the classical index $i$ is suppressed.
Note that even though $a$ is exponentially small, the individual amplitudes of this state are of similar size. Reducing the state in terms of the $k$ index leads to
 \begin{eqnarray}
{\rm tr}_2 ( |\chi\rangle\langle \chi| ) = 
\frac{1}{C}  \biggl(\sum_{j,j'} |j\rangle  \langle j' |  \sum_{k=1}^{N/2} \Big( F_{j,k} |0\rangle  + 
 a (F^\dagger F)_{j,k} |1\rangle \Big)  \Big(F_{j',k}^\ast  \langle 0|   + 
 a(F^\dagger F)_{j',k}^\ast \langle 1| \Big) \biggr) .\nonumber
 \end{eqnarray}	
In matrix form, this reduced density matrix is written as
 \begin{equation}
 G:= \frac{1}{C}\left[
 \begin{array}{cc}
 F F^\dagger & a\, F (F^\dagger F) \\
 a\, (F^\dagger F) F^\dagger & a^2\,(F^\dagger F)(F^\dagger F)
 \end{array}
 \right].
 \end{equation}
 By the use of the singular value decomposition of $F=US V^\dagger$, this matrix---positive semi-definite by construction---can 
 be written as
\begin{equation}
	 G=\frac{1}{C}\left[
 \begin{array}{cc}
U &0  \\
  0 & V
 \end{array}
 \right] \left[
 \begin{array}{cc}
 S^2 & a\,S^3\\
 a\,S^3 & a^2 S^4
 \end{array}
 \right]
 \left[
 \begin{array}{cc}
 U^\dagger   &0  \\
  0 & V^\dagger 
 \end{array}
 \right] .
\end{equation}  
In precisely the same way, we are given multiple copies of the state 
 \begin{equation}
|\widetilde{\chi}\rangle = \frac{1}{\sqrt{C}} \sum_{j,k=1}^{N/2} |j\rangle |k\rangle \left( a (FF^\dagger )_{j,k}  |0\rangle + 
F_{j,k}^\dagger |1\rangle\right).
 \end{equation}	
 Again reducing the state in terms of the $k$ index leads to
 \begin{eqnarray}
{\rm tr}_2 ( |\widetilde{\chi}\rangle\langle \widetilde{\chi}| ) = 
\frac{1}{C}  \biggl(\sum_{j,j'} |j\rangle \langle j' | \sum_{k=1}^{N/2} \Big(  a (F F^\dagger)_{j,k}   |0\rangle  + F_{j,k}^\dagger  |1\rangle \Big)  \Big( a(F F^\dagger)_{j',k}^\ast \langle 0|   + (F_{j',k}^\dagger)^\ast
   \langle 1| \Big) \biggr),\nonumber
 \end{eqnarray}	
leading to the matrix
 \begin{equation}
 \widetilde{G}:= \frac{1}{C}\left[
 \begin{array}{cc}
 a^2\,(F F^\dagger)(FF^\dagger ) & a\,   (F F^\dagger ) F \\
 a\, F^\dagger (F F^\dagger ) &  F^\dagger F
 \end{array}
 \right],
 \end{equation}
 which can be decomposed as
\begin{equation}
	 \widetilde{G}=\frac{1}{C}\left[
 \begin{array}{cc}
U &0  \\
  0 & V
 \end{array}
 \right] \left[
 \begin{array}{cc}
 a^2 S^4 & a\,S^3\\
 a\,S^3 & S^2
 \end{array}
 \right]
 \left[
 \begin{array}{cc}
 U^\dagger   &0  \\
  0 & V^\dagger 
 \end{array}
 \right].
\end{equation}  
The matrix 
\begin{equation}
	Z:= \frac{1}{2}(G+\widetilde{G})
\end{equation}
has still low rank, as it has just twice the rank of $F$. Its eigenvectors are $(u_j,\pm v_j)\in \CC^N$ and its eigenvalues in terms of the singular values of $F$ are given by 
$s_j^2(a s_j\pm 1)^2/(2C)$ since
\begin{equation}
Z=\frac{1}{2C}\left[
\begin{array}{cc}
F F^\dagger + a^2\,(F F^\dagger)(FF^\dagger )  & 2 a\, F F^\dagger F \\
2 a\, F^\dagger F F^\dagger  & a^2\,F^\dagger F + (F^\dagger F)(F^\dagger F)
\end{array}
\right] 
\end{equation}
and
\begin{align}
\frac{1}{2C}\left[
\begin{array}{cc}
F F^\dagger + a^2\,(F F^\dagger)(FF^\dagger )  & 2 a\, F F^\dagger F\\
2 a\, F^\dagger F F^\dagger  & a^2\,F^\dagger F + (F^\dagger F)(F^\dagger F)
\end{array}\nonumber
\right] &\left[ \begin{array}{c} u_j \\ \pm v_j \end{array}\right] \\=  \frac{1}{2C} \left[ \begin{array}{c} (s_j^2+ a^2 s_j^4 \pm 2 a s_j^3 )u_j \\ (2 a s_j^3 \pm s_j^2 \pm a^2 s_j^4 ) v_j \end{array}\right] =  \frac{1}{2C}s_j^2(a s_j\pm 1)^2 &\left[ \begin{array}{c} u_j \\ \pm v_j \end{array}\right]. 
\end{align}
This renders standard quantum principal component analysis~\cite{Lloyd2014} readily applicable.

 \end{document}